\documentclass[preprint,authoryear,review,12pt]{elsarticle}

\usepackage{amsmath, amssymb, amscd, amsthm}
\usepackage{amsfonts}
\usepackage{bbm}
\usepackage[all]{xy}
\usepackage{graphics}
\usepackage{graphicx}
\usepackage{epsfig}  
\usepackage{latexsym}

\theoremstyle{plain}

\newcommand{\LL}{\mathcal{L}}
\newcommand{\R}{\mathcal{R}}
\newcommand{\C}{\mathcal{C}}

\newcommand{\E}{\mathcal{E}}
\newcommand{\G}{\mathcal{G}}
\newcommand{\X}{\mathcal{P}}
\newcommand{\qq}{Q}
\newcommand{\qzero}{q_0}
\newcommand{\qone}{q_1}
\newcommand{\qtwo}{q_2}
\newcommand{\vectorq}{\vec{q}}

\newcommand{\set}[1]{\{ #1 \}}

\newcommand{\del}{\partial}

\journal{Journal of Theoretical Biology}

\begin{document}

\title{Higher order Boolean networks\\ as models of cell state dynamics}

\begin{frontmatter}

\author{Elke K Markert$^1$, Nils Baas$^2$, Arnold J. Levine$^{1,3}$ and 
Alexei Vazquez$^{1,3,4}$\corref{cor1}}

\address{$^1$ Simons Center for Systems Biology, Institute for Advanced 
Study, 1 Einstein Dr, Princeton, NJ 08540, USA}

\address{$^2$ Department of Mathematics, Norwegian University of Science 
and Technology, Trondheim, Norway}

\address{$^3$ The Cancer Institute of New Jersey, 195 Little Albany St, 
New Brunswick, NJ 08963, USA}

\address{$^4$ Department of Radiation Oncology, UMDNJ-Robert Wood Johnson 
Medical School}

\cortext[cor1]{Corresponding author}

\begin{abstract}

The regulation of the cell state is a complex process involving several 
components. These complex dynamics can be modeled using Boolean networks, 
allowing us to explain the existence of different cell states and the 
transition between them. Boolean models have been introduced both as 
specific examples and as ensemble or distribution network models. 
However, current ensemble Boolean network models do not make a systematic 
distinction between different cell components such as epigenetic factors, 
gene and transcription factors. Consequently, we still do not understand 
their relative contributions in controlling the cell fate. In this work we 
introduce and study higher order Boolean networks, which feature an 
explicit distinction between the different cell components and the types 
of interactions between them. We show that the stability of the cell state 
dynamics can be determined solving the eigenvalue problem of a matrix 
representing the regulatory interactions and their strengths. The 
qualitative analysis of this problem indicates that, in addition to the 
classification into stable and chaotic regimes, the cell state can be 
simple or complex depending on whether it can be deduced from the 
independent study of its components or not. Finally, we illustrate how the 
model can be expanded considering higher levels and higher order dynamics.

\end{abstract}

\begin{keyword}
cellular automata \sep complex systems \sep epigenetics \sep transcription 
\sep protein modifications
\end{keyword}

\end{frontmatter}

\bibliographystyle{elsarticle-harv}

\section{Introduction}

Regulation of gene expression is a complex process involving several 
components of different type, such as epigenetic factors, gene and 
transcription factors. Modeling such a complex system requires us to find 
the balance between the accuracy of the model predictions and our ability 
to interpret the model. On one side of the model spectrum, we have {\em 
detailed} chemical kinetics or Boolean network models \cite{jong02}. In 
these approaches the cell component heterogeneity is build in when 
specifying the regulatory interactions, functions (kinetic models or 
Boolean functions), and associated model parameters. Provided we determine 
all the regulatory interactions, functions and parameters correctly, these 
models can allow us to make accurate predictions of the cell state 
dynamics. However, detailed models can be queried only by means of 
numerical simulations, making it difficult to uncover or understand any 
behavior that is not known in advance. On the other end of the model 
spectrum we have {\em ensemble} models, which specify the statistical 
distributions of the regulatory interactions, functions, and associated 
model parameters. While these models cannot provide precise predictions 
about specific cell processes, they can allow us to understand what is the 
typical behavior and how it can change under variation of the model 
parameters. Within this class of models, ensemble Boolean networks have 
been studied the most 
\cite{kauffman69,derrida86,kauffman93,aldana03,kauffman04}.

The analysis of ensemble Boolean networks has significantly contributed to 
our qualitative understanding of the cell state dynamics 
\cite{kauffman69,kauffman93}. Different cell states can be associated with 
different stable attractors of the Boolean network dynamics 
\cite{kauffman69,kauffman93} and we can study the breakdown of this 
stability following parameter changes 
\cite{derrida86,aldana03,kauffman04}. More recently it is becoming clear 
that not all transcription factors regulating a given gene are equivalent. 
This is being modeled using Boolean functions with a biologically 
meaningful structure, such as canalyzing functions 
\cite{harris02a,kauffman04} and nested canalyzing functions 
\cite{kauffman04}. However, at the system level, the current ensemble 
Boolean network models typically comprise all elements they consider into 
one class of objects, within which the interactions are determined. This 
makes it difficult to model the general behavior and influence of 
different groups of elements (cell components) and the different types of 
interactions which systematically occur between elements of these 
components.

We introduce a more general class of Boolean networks with an explicit 
distinction between epigenetic factors, genes and transcription factors 
and the types of interactions among them. We call this class of Boolean 
networks higher order Boolean networks (HOBN), in the sense that we 
specify wiring diagrams both within the three groups and between them, 
determining type-level interactions. We use HOBN to investigate the 
relative contribution of the different cell components to the cell state 
dynamics.

\section{Higher order Boolean network model}

\begin{figure}

\centerline{\includegraphics[width=3.3in]{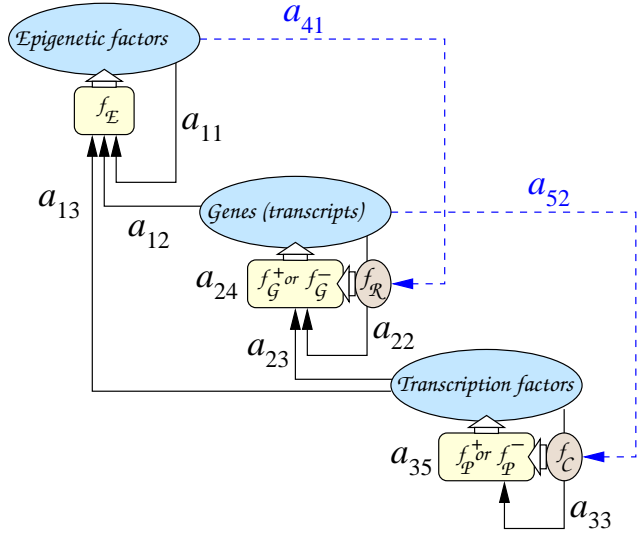}}

\caption{Higher order representation of the cell components and their 
regulatory interactions. The square boxes indicate the sets of Boolean 
functions controlling the states of the elements, the arrows entering a 
box indicate the input of these functions. In the case of gene expression, 
the choice of function is regulated by epigenetic state; in the case of 
transcription factors, the choice of function is regulated by expression 
of components (AND-function).}

\label{figure:regulationgraph} 

\end{figure}

We model the interaction of three different types of cellular components 
determining the cell state (Fig. \ref{figure:regulationgraph}): a set of 
epigenetic factors $\E$, a set of genes or transcripts $\G$, and finally a 
set of transcription factors or proteins $\X$.

Epigenetic factors form the most basic elements of our system, 
representing chemical modifications of the DNA and histone tails 
\cite{jaenisch03}. The list of all such factors for a cell genome is the 
set $\E$. They are distributed along the DNA forming a linear graph. 
Within this topological graph they have neighbors influencing their 
states. Each epigenetic factor $e\in\E$ thus comes with a neighborhood 
$\E_e$ of other elements in $\E$, describing a linear order. The elements 
in this neighborhood influence the epigenetic state of the system. 
Epigenetics can be influenced also by single and composite gene products 
which can alter for example methylation patterns. We model this by 
assigning to each factor $e\in\E$ not only its direct neighbors within 
$\E$, but also elements in the set of genes $\G$ and in the set of 
transcription factors $\X$. Thus an element $e\in\E$ has three 
neighborhoods $\E_e$, $\G_e$ and $\X_e$ which regulate its state. The 
epigenetic factors are assumed to be in two possible states 0 (e.g. not 
methylated) and 1 (e.g. methylated). We model the control of the {\it 
epigenetic state} by a distribution of Boolean functions $f_\E$ on the set 
$\E$ which each take as input the states of all three neighborhood 
elements of a given epigenetic factor, and give as output the updated 
state 0 or 1. We think of epigenetics as the most elementary units of the 
system since these factors act upon genes (and consequently on 
transcription factors) in a dominant fashion: an adequate epigenetic state 
is a {\it prerequisite} of all other transcription and translation 
activities, including up- or down-regulation of transcription by various 
factors.

A gene or transcript $g\in\G$ represents any genomic region that can be 
transcribed, including mRNAs of genes, miRNAs and short RNAs. The active 
state indicates the gene presence. The state of a gene is regulated by 
epigenetic factors, other genes and transcription factors, however in 
different ways. Epigenetic factors determine the secondary DNA structure 
in their neighborhood and the accessibility of this region to 
transcription factors and the transcription machinery. Thus the aggregate 
epigenetic state of these factors on and nearby the DNA segment encoding 
for a gene can influence the gene's transcription rate, and thus the gene 
state. We model this by introducing two different transcription regimes 
characterized by two Boolean functions $f_\G^-$ and $f_\G^+$, where 
$f_\G^-$ corresponds to the silent or restricted regime and $f_\G^+$ 
corresponds to the active or accessible regime. To model the epigenetic 
regulation of the transcription regime of a gene we define the set of all 
possible transcription regimes $\R$ (here, $\R=\set{f_\G^+, f_\G^-}$), 
together with an additional Boolean function $f_{\R}$ which controls the 
{\em change of transcription regimes}. This function $f_\R$ takes as input 
the state of the epigenetic factors associated to a gene $\E_g$ and the 
value of its present regime ($f_\G^-$ or $f_\G^+$) and determines as an 
output the value of the regime for the next step. Once a regime is 
determined depending on the epigenetic state of the gene, the regime 
function will update the gene state. This update now happens depending on 
gene-gene interactions and transcription factor activities. Thus the 
inputs of the Boolean functions in $\R$ are taken from the set of genes 
and transcription factors which regulate transcription of a gene. This 
means that each gene comes also with a neighborhood of other genes and a 
neighborhood of transcription factors regulating its state. The gene 
neighborhood of a given gene $g$ will be called $\G_g$ and we denote by 
$\X_g$ its transcription factor neighborhood. The state of $g$ is now 
regulated by the states of elements in $\G_g$ and $\X_g$ through a 
transcription regime, $f_\G^-$ or $f_\G^+$, the choice of which is decided 
by the epigenetic state $\E_g$ of the gene. The model thus allows us to 
control epigenetic effects separately from the regulation by other 
transcripts and transcription factors.

The last group of cellular agents consists of the set of transcription 
factors $\X$, representing proteins and protein complexes. Each element 
$p$ in $\X$ is composed of products of a subset of genes $\G_p$. In order 
for a transcription factor $p$ to be assembled, all the transcripts in 
$\G_p$ need to be transcribed and translated. This procedure - which 
enables or disables the activity of transcription factors - is again best 
modeled by a regime switch. We assume the transcription factors can be in 
two different regimes characterized by the Boolean functions $f_\X^-$ and 
$f_\X^+$, where $f_\X^-\equiv 0$ corresponds to the not assembled complex 
and $f_\X^+$ to the regime of the assembled complex. The set of all 
regimes for protein complexes is denoted by $\C$. The choice of regime for 
a protein $p$ will depend on the states of all elements in $\G_p$ via a 
Boolean function $f_\C$. This regime switch function $f_\C$ is simply a 
logical $AND$ relation, since the transcription factor works under 
$f_\X^+$ if and only if all its components are transcribed, i.e. if and 
only if all inputs of $f_\C$ are in $1$-state. Here the $1$-state of 
$f_\C$ stands for $f_\X^+$ while $0$ stands for $f_\X^-$. Within the 
positive regime $f_\X^+$ the state of an element in $\X$ can depend on 
interaction with various other elements in $\X$ itself; for example via 
post-translational modifications \cite{walsh05}. These form the 
neighborhood $\X_p$ of $p\in\X$. The functions in the regime $f_\X^+$ thus 
take as input the states of elements in this neighborhood.

The cellular components and regulatory interactions just described are 
summarized in the wiring diagram shown in Fig. 
\ref{figure:regulationgraph}. This diagram emphasizes the level-wise 
organization of the different types of cellular components. From top to 
bottom we have {\em regime} regulation (dashed lines). From bottom to top 
we have {\em state} regulation, as considered by previous Boolean network 
models (solid lines). When both types of interactions are put together we 
obtain a closed system, a system with feedback, which is a distinctive 
property of cell regulatory networks \cite{barabasi04}. Following the 
central dogma, we will often refer to epigenetics, i.e. the set $\E$, its 
elements, neighborhoods and Boolean rule, as the $0$-level of the system. 
The set $\G$ together with the two Boolean rules, neighborhoods, and 
associated epigenetic factors constitute the $1$-level, while finally the 
set of transcription factors $\X$ together with its own functions, 
neighborhoods and associated genes will be called the $2$-level. We will 
also sometimes refer to the interactions regulating state changes as {\em 
primary interactions}, while those regulating a choice of regimes will be 
called {\em secondary interactions}. This is also in reference to the 
existing notion of higher order cellular automata, as introduced by 
\cite{baas05}.

\section{Cell state dynamics}

Previous studies of Boolean network models indicate the existence of two 
dynamical modes. An {\em ordered mode} where two different trajectories in 
the cell state space will converge to the same trajectory, and a {\em 
chaotic mode} where the trajectories will instead diverge 
\cite{derrida86}. Later on it was shown that the ordered mode implies a 
nearly static system behavior where most elements ({\it stable core}) are 
not changing state \cite{flyvbjerg88}. Here, we follow the latter 
approach. The total state of our system is expressed by {\it five state 
variables}: the epigenetic state $e(t)$ for $e\in\E$, the gene state 
$g(t)$ for $g\in\G$, the transcription factor state $p(t)$ for $p\in\X$, 
and furthermore the transcription regime state $r(t)$ and the protein 
regime state $c(t)$, where in the latter two cases the $-$ and $+$ states 
are represented by 0 and 1 respectively. The basic set-up for the system 
dynamics thus consists of five equations expressing the probability of 
changes to happen in all of the five state variables. We can think of the 
total probability of change of state of the system as a five-dimensional 
vector $\vectorq(t) = (\qzero(t), \qone(t), \qtwo(t), \qq_1(t), 
\qq_2(t))^t$, where $\qzero(t)$, $\qone(t)$, $\qtwo(t)$, $\qq_1(t)$ and 
$\qq_2(t)$ are the probability that a given epigenetic, gene, 
transcription factor, transcription regime, and protein regime state 
respectively, will change from step $t$ to step $t+1$.

In the general case $\vectorq(t+1)$ is a nonlinear function of 
$\vectorq(t)$, which depends on the detailed definition of the Boolean 
model. Nevertheless, in the nearly static, ordered mode, where most 
elements do not change state, this function can be linearized in good 
approximation: in this range the absolute value of the total probability 
for change in the system is very close to $0$, i.e. $|\vectorq| 
\rightarrow 0$, which allows us to neglect higher exponent terms. Note 
that this linear approximation of the system dynamics breaks down outside 
the ordered mode and cannot predict any behavior there, other than the 
fact that the system is in an unstable mode. In the near-static range we 
thus obtain

\begin{equation}\label{qtt}
\vectorq(t+1) = A \vectorq(t)\ ,
\end{equation}

\noindent where $A$ is a five by five positive definite matrix. The 
entries in the matrix $A$ reflect the regulatory patterns indicated in 
Fig. \ref{figure:regulationgraph} and therefore it is of the form

\begin{equation}\label{A}
A = \left( \begin{array}{ccc|cc}
a_{11} & a_{12} & a_{13} & 0 & 0\\
0 & a_{22} & a_{23} & a_{24} & 0\\
0 & 0 & a_{33} & 0 & a_{35}\\
\hline
a_{41} & 0 & 0 & 0 & 0 \\
0 & a_{52} & 0 & 0 & 0 \end{array} \right)
\end{equation}

\noindent Each set in Fig. \ref{figure:regulationgraph} which is acted on 
or regulated creates one state dimension in $\vec{q}$. Each arrow 
indicating regulation is represented by one non-trivial entry in the 
matrix. Notice that the regime control interactions (dashed lines) are 
represented in the off-diagonal blocks, while the state control, or 
ordinary interactions (solid lines) appear in the upper triangular 
submatrix.

In the following we focus on the stability of the linear map (\ref{qtt}). 
Specifically, the map is said to be {\em stable} when 
$|\vec{q}|\rightarrow 0$ as $t\rightarrow\infty$, and it is called {\em 
unstable} otherwise. The stability of a linear map can be deduced from the 
properties of the eigenvalues of the corresponding matrix, in this case 
$A$. When the largest eigenvalue has absolute value less than one the 
system is in stable or ordered mode and $\vectorq(t)$ converges to zero. 
If however the largest eigenvalue becomes larger than one, this indicates 
that the system is in unstable mode where the linearization (\ref{qtt}) is 
not a suitable approximation to the actual system anymore. Using the 
linearization (\ref{qtt}), we can thus analyze the model within its {\em 
`stable range'}, i.e. in its near static mode. We can furthermore 
determine the conditions on parameter space which distinguish between 
stable and unstable range, by calculating the largest eigenvalue and 
setting it equal to one. Finally, we can analyze the derivative in time 
direction to determine the influence of certain parameters on the growth 
of the largest eigenvalue.

Since $A$ is derived from a strongly connected graph (see Fig. 
\ref{figure:regulationgraph}), $A$ is {\it irreducible}. Furthermore the 
entries of $A$ are non-negative. In this case we can apply the Frobenius 
Theorem for non-negative irreducible matrices \cite{gantmacher00}. This 
theorem guarantees that $A$ has a real positive eigenvalue $\Lambda$ such 
that any other eigenvalue $\lambda$ of $A$ satisfies $|\lambda|\le 
\Lambda$.

\begin{figure}

\centerline{\includegraphics{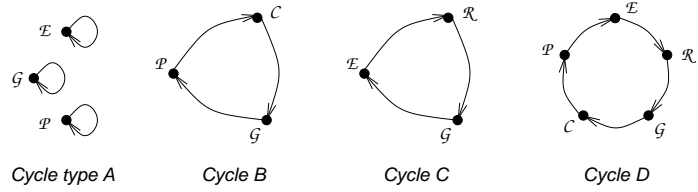}}

\caption{Effective dynamic modes}

\label{figure:cycles}
\end{figure}

The largest eigenvalue of $A$ can be obtained finding the roots of the 
characteristic polynomial $P(\lambda)=\det(\lambda I -A)$, where $I$ is 
the identity matrix. In our case $P(\lambda)$ is given by the quintic 
polynomia

\begin{eqnarray}\label{l4}
P(\lambda)
 &=& \lambda^2(A_{\E}-\lambda)(A_{\G}-\lambda)(A_{\X}-\lambda)
\nonumber\\
&-& \lambda(A_{\E}-\lambda)B - \lambda(A_{\X}-\lambda)C + D
\end{eqnarray}

\noindent where

\begin{eqnarray}\label{AD}
A_{\E} = a_{11}\ \ \ \ A_{\G}= a_{22}\ \ \ \ A_{\X} = a_{33}\ ,\nonumber\\
B = a_{23}a_{35}a_{52}\ \ \ \ C = a_{24}a_{41}a_{12}\ ,\nonumber\\
D = a_{24}a_{41}a_{13}a_{35}a_{52}\ .
\end{eqnarray}

\noindent The direct inspection of this equation tells us about the basic 
dynamic cycles of the system (Fig. \ref{figure:cycles}). There are three 
minimal self controlling cycles, represented by the type A cycles in Fig. 
\ref{figure:cycles}. The gene self-cycle represents the standard type of 
dynamics in ordinary Boolean network models, where genes are regulated by 
other genes in a regulation network. This is demonstrated in more detail 
in Section \ref{section:examples}. Similarly, the epigenetics self-cycle 
models how epigenetic factors update their states based on their previous 
states and the states of their neighbors along a DNA segment. The 
transcription factor self-cycle models changes in the transcription 
factors state due to interactions between them and therefore represents 
regulation at the post-translational level. There are two cycles composed 
of three components, represented by the cycles B and C in Fig. 
\ref{figure:cycles}. $B$ represents a dynamic mode where a change in gene 
state alters the protein regime state and thus the state of transcription 
factors. This in turn alters genes states, while epigenetic factors and 
the transcription regime remain unvariable. $C$ involves state changes in 
epigenetic factors, transcription regime and genes. Finally, cycle D is 
composed of five elements, representing a dynamic mode where changes in 
the epigenetic states result in changes of transcription regime, leading 
to changes in the genes states, thus altering the transcription factor 
regime and transcription factor states, which then go on and change the 
epigenetic factor states.

Although we cannot explicitly solve the quintic polynomial equation 
determining the roots of (\ref{l4}), we can derive some general results. 
For example, the boundary separating the stable from the unstable regime 
is given by the equation $P( 1; A_\E, A_\G, A_\X, B, C, D) = 0$. We 
investigate also the relative influence of the different cycles 
calculating the partial derivatives of $\Lambda$ with respect to the 
effective coefficients in (\ref{AD}) (see Supp. Material). It can be shown 
that in the stable range ($\Lambda<1$), $\Lambda$ is most sensitive in the 
direction of $D$. In other words the easiest way to make a transition to 
the unstable regime is to increase $D$. This result indicates that the 
dynamic mode $D$, which makes full use of all regulatory mechanisms, 
dominates over the other modes. Consequently all regulatory mechanisms are 
coupled together and their influence in the cell state dynamics cannot be 
analyzed independently of each other. This type of structural analysis can 
provide insight into the regulation of the system as a whole without even 
breaking the calculations down to the actual ``microscopic parameters'' 
characterizing the neighborhoods, memberships and the Boolean functions. 
Furthermore, provided we have quantitative estimate of the model 
parameters we can always numerically compute the largest eigenvalue of $A$ 
and determine whether the system is or is not in an stable regime.

\section{\bf Neighborhoods, memberships, Boolean functions and updating 
schemes}

The matrix elements of $A$ can be derived from the properties of 
neighborhoods, memberships and Boolean functions. In this way we can also 
investigate the influence of ``microscopic'' parameters on the cell state 
dynamics. For example, let us assume that, given a type of elements and 
neighborhoods, all neighborhoods have the same size (in an ensemble 
network one will use the estimated mean value). In this case there are 
three neighborhood parameters $K_0^0$, $K_0^1$ and $K_0^2$ for the three 
types of neighborhoods ($\E_e$, $\G_e$, $\X_e$) on the $0$-level 
(epigenetics). On the $1$-level (genes) there are two of those, $K_1^1$ 
and $K_1^2$, and finally on the $2$-level (transcription factors) there is 
only one $K_2^2$. These parameters also give the input lengths for the 
Boolean rules determining change of state on the three levels. Furthermore 
the change of regime rule $f_\R$ on the $1$-level has input length $M_0$, 
the number of ``members'' constituting the epigenetic state of a gene. 
Finally, the rule $f_\C$ changing transcription factor regime on the 
second level has input length $M_1$, the number of transcription factor 
members.

The Boolean functions can be characterized by the probability that two 
different inputs result in a different output (sensitivity 
\cite{shmulevich04}) and the probability that the output is 1 for a 
randomly chosen input. We assume the Boolean functions $f_\E$, 
$f^{\pm}_\G$, $f_\X^+$, and $f_{\R}$ are randomly sampled from the 
function classes $F_\E$, $F_\G$, $F_\X$, and $F_{\R}$ respectively, while 
$f_\X^-\equiv0$ and $f_{\C}\equiv AND$. These function classes are 
characterized by their expected sensitivity $s_\E$, $s^{\pm}_\G$, 
$s_\X^-=0$, $s^+_\X$, $s_{\R}$ and $s_{\C}$ and probability to be in the 
$1$-state $\rho_\E$, $\rho_\G^{\pm}$, $\rho_\X^-=0$, $\rho_\X^+$, 
$\rho_{\R}$ and $\rho_{\C}$, respectively. The expected sensitivities 
depend on the class of Boolean function \cite{shmulevich04}. For example, 
when the functions are sampled from a distribution with a given $\rho$ we 
have $s=2\rho(1-\rho)$. On the other hand, for $f_{\C}=AND$ we obtain 
$\rho_{\C}=\rho_\G^{M_1}$ and $s_{\C}=\rho_\G^{M_1-1}$, where $\rho_\G = 
\rho_\R\rho_\G^+ + (1-\rho_\R)\rho_\G^-$ is the probability that a given 
input of the $AND$-rule is in state $1$. Finally, the matrix elements will 
also depend on the specific updating scheme. In the following we will 
assume a synchronous updating scheme, where the state of all elements in 
the systems are updated simultaneously.

Overall, we have a system with $15$ parameters. The coefficients $a_{ij}$ 
can be derived from these parameters. Most of them can be determined as in 
previous Boolean network models, consisting of the product of the 
sensitivity and the neighborhood size. This is the case for $a_{1i} = s_\E 
K_0^{i-1}$ ($i=1,2,3$), $a_{2i} = s_\G K_1^{i-1}$ ($i=1,2$), and $a_{33} = 
s_\X K_2^2$, where $ s_\G = \rho_{\R} s_\G^+ + (1-\rho_{\R}) s_\G^-$ and $ 
s_\X = \rho_{\C} s_\X^+$ are the expected sensitivities of the respective 
Boolean functions after accounting for regime changes. The coefficients 
characterizing the change in regimes can be calculated in a similar way, 
but replacing neighborhood sizes by membership sizes. This is the case for 
$a_{41} = s_{\R} M_0$ and $a_{52} = s_{\C} M_1$. Finally, the coefficients 
representing state changes following regime changes are calculated 
differently. What matters in this case is the probability that two Boolean 
functions from different regimes result in a different output given the 
same input. This is the case for $a_{24}= s(f_\G^- \leftrightarrow 
f_\G^+)= \rho_\G^+(1-\rho_\G^-)+\rho_\G^-(1-\rho_\G^+)$ and $a_{35} = 
s(f_\X^- \leftrightarrow f_\X^+)= 
\rho_\X^+(1-\rho_\X^-)+\rho_\X^-(1-\rho_\X^+)$. Taking these results 
together we obtain

\begin{eqnarray}\label{coefficients}
a_{11} &=& s_{\cal E} K_0^0\ ,\ \ \ \ 
a_{12} = s_{\cal E} K_0^1\ ,\ \ \ \ 
a_{13} = s_{\cal E} K_0^2\ ,
\nonumber\\
a_{22} &=& s_{\cal G} K_1^1\ ,\ \ \ \
a_{23} = s_{\cal G} K_1^2\ ,\ \ \ \
a_{24} = \rho_\G^+(1-\rho_\G^-)+\rho_\G^-(1-\rho_\G^+)\ ,
\nonumber\\
a_{33} &=& s_{\cal P} \rho_\G^{M_1} K_2^2\ ,\ \ \ \
a_{35} = \rho_{\cal P}^+
\nonumber\\
a_{41} &=& s_{\cal R}M_0\ ,\ \ \ \
a_{52} = M_1\rho_\G^{M_1-1}
\end{eqnarray}

From the Frobenius Theorem it follows that the largest eigenvalue 
$\Lambda$ is a monotonic increasing function of the matrix elements 
$a_{ij}$ \cite{gantmacher00}. Therefore we can always investigate the 
contribution of any of the parameters listed above by analyzing their 
influence on the matrix elements $a_{ij}$. For fixed sensitivities, the 
increase of any neighborhood size $K_i^j$ always result in the increase of 
at least one matrix element, pushing the system towards the unstable 
regime. In contrast, the number of members of a transcription factor $M_1$ 
influences $a_{33}$ and $a_{52}$ (\ref{coefficients}), which are both 
decreasing functions of $M_1$ provided $M_1\geq1$. Therefore, the larger 
the protein complexes are, the more stable is the system.

The Frobenius Theorem also provides bounds for the largest eigenvalue, 
namely by the minimum and maximum of the row sums of the matrix $A$: 
$\min_i(\sum_ja_{ij})\leq\Lambda\leq\max_i(\sum_ja_{ij})$ 
\cite{gantmacher00}. In the current example we obtain

\begin{equation}\label{bounds}
\sum_ja_{ij} = \left\{
\begin{array}{ll}
s_\E(K_0^1+K_0^1+K_0^2)\ , & i=1\\
s_\G(K_1^1+K_1^2) + 
s(f_\G^- \leftrightarrow f_\G^+) \ & i=2\\
s_\X \rho_\G^{M_1}K_2^2 +
s(f_\X^- \leftrightarrow f_\X^+)\ , & i=3\\
s_\R M_0\ , & i=4\\
M_1\rho_\G^{M_1-1}\ , & i=5
\end{array}
\right.
\end{equation}

\noindent These sums put together all the contributions regulating the 
units states at each level. When all of them are smaller (or larger) than 
1 we can guarantee that $\Lambda<1$ ($\Lambda>1$) and the system is in a 
stable state (or unstable state, respectively). However, when some are 
smaller and other are larger than 1, we are forced to compute $\Lambda$ to 
determine the system stability. In other words, the system exhibits 
non-trivial {\it complex} behavior, where fundamental properties of the 
system as a whole are not directly coupled with the corresponding 
properties of its subsystems.

\begin{figure}

\centerline{\includegraphics[width=5in]{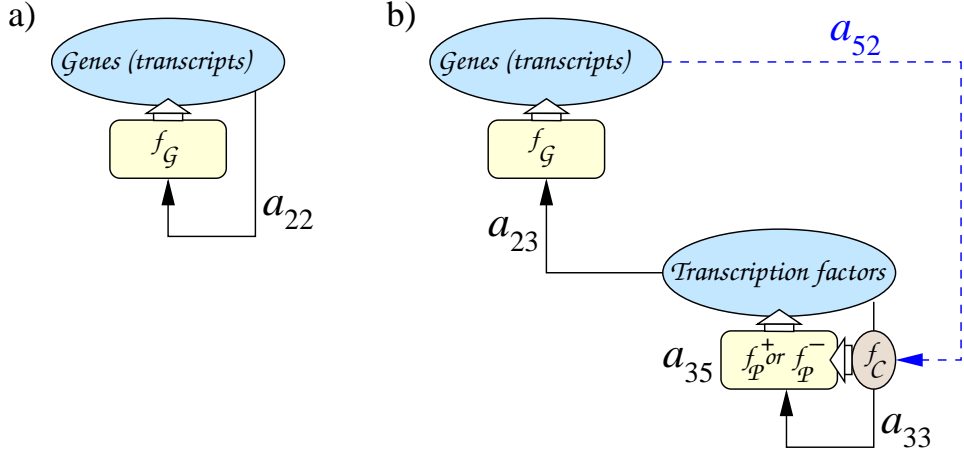}}

\caption{{\bf Reduced versions of the HOBN depicted in Fig. 
\ref{figure:regulationgraph}} a) Standard Boolean network model. b) 
Boolean network model accounting for the formation of protein complexes 
and regulation at the protein level.}

\label{fig:example}

\end{figure}

\section{Examples}\label{section:examples}

To illustrate the concepts introduced above we discuss a few examples, 
allowing us to emphasize the flexibility of this modeling framework and 
the importance of including the regulatory structure at the level of 
components.

{\it Standard Boolean network:} Here we show how we can reduce our model 
to compare directly to the standard Boolean network models considered in 
the literature so far, with the reduced wiring diagram shown in Fig. 
\ref{fig:example}a. In this case, there is no regulation at the epigenetic 
level and at the transcription factor level, nor at the transcription 
regime level. The only dynamic mode is the self-cycle A$_\G$ in Fig. 
\ref{figure:cycles} and the characteristic polynomial (\ref{l4}) is 
reduced to $P(\lambda) = \lambda^4 (\lambda-A_\G)$. In this case, the 
largest eigenvalue is $\Lambda=A_\G$ and therefore the effective parameter 
controlling stability is $\theta=A_\G$. In particular, taking into account 
that $A_\G=a_{22}$ (\ref{AD}), and assuming constant neighborhood sizes 
and a synchronous update (\ref{coefficients}), we obtain

\begin{equation}\label{thetaG}
\theta=s_{\cal G}K_1^1.
\end{equation}

\noindent This is precisely the result obtained for the classical standard 
Boolean network model \cite{kauffman69,kauffman93}. Thus, our approach 
contains the standard Boolean network model as a special case.

{\it Regulation at the protein level:} In the next example we study 
regulation on the transcription factor level Fig. \ref{fig:example}b. As 
opposed to the first example, here we have proteins and protein compounds 
regulating transcription; this includes the assembly rule, i.e. the regime 
change on transcription factor level. The only non-zero matrix elements 
are those corresponding to arrows in Fig. \ref{fig:example}b and the only 
relevant dynamic modes are the cycles A$_{\cal P}$ and B in Fig. 
\ref{figure:cycles}. The characteristic polynomial ({\ref{l4}) now reduces 
to $P(\lambda) = -\lambda^2 ( \lambda^3- A_{\cal P} \lambda^2 - B )$. 
While finding the roots of this cubic polynomial can be cumbersome, 
finding the stability condition is in this case straigthforward. At 
$\lambda=1$ we obtain the stability condition $B+A_{\cal P} = 1$. 
Furthermore, since the largest eigenvalue $\Lambda$ of $A$, $A_{\cal P}$ 
and $B$ are all continuous increasing functions of the associated matrix 
elements of $A$, then $\Lambda<1$ for $B+A_{\cal P}<1$ and $\Lambda>1$ 
when $B+A_{\cal P} > 1$. So in this case, the effective control parameter 
for stability is given by $\theta = B + A_{\cal P}$. In particular, 
assuming constant neighborhood and membership sizes and synchronous 
updates, from (\ref{AD}) and (\ref{coefficients}) we obtain

\begin{equation}\label{thetaP} 
\theta = s_{\cal G} K_1^2 \rho_{\X}^+ M_1\rho_{\cal G}^{M_1-1} 
+ s_\X K_2^2 \rho_\G^{M_1}
\end{equation}

\noindent Notice that this formula consistently solves the following 
subtlety: instead of speaking of gene-gene regulation networks as in the 
previous example, one could actually distinguish between the gene and its 
products (proteins), and construct a network where single gene products 
regulate gene transcription. This would be a more accurate description of 
the biological reality, should, however, lead to the same results, since 
de facto we do not change any interactions. In our model, this set-up 
would mean that we set $M_1=1$, i.e. all protein complexes are single gene 
products; furthermore $K_2^2=0$, i.e. there is no protein-protein 
regulation (only proteins regulating genes), and lastly $K_1^2=K_1^1$, 
since we have replaced the number of neighbor genes regulating a gene by 
the same number of their proteins. We obtain $\theta=s_{\cal G} 
K_1^2=s_{\cal G} K_1^1$, which coincides perfectly with (\ref{thetaG}). 
Therefore, unless we account for the formation of protein complexes or 
some regulation at the protein level, we obtain the same effective control 
parameter of the standard Boolean network model.

The addition of protein-protein interactions results in the second term in 
the r.h.s. of (\ref{thetaP}). Here - just as for the case of gene 
regulation - increasing the neighborhood size $K_2^2$ increases $\theta$, 
pushing the system towards the unstable mode. On the other hand, $\theta$ 
decreases exponentially with increasing the protein complex size $M_1$, 
making the system more stable. We see from the above formulas how the 
stability conditions change drastically depending on primary and secondary 
interaction parameters. These corrections emphasize the importance of 
considering the right structure in the modeling framework. Furthermore, 
although there is an increase in model complexity, we can still derive 
analytical results allowing us to obtain a better qualitative 
understanding of the cell state dynamics.

\section{Beyond three levels}\label{section:generalities2}

\begin{figure}

\centerline{\includegraphics[width=3.2in]{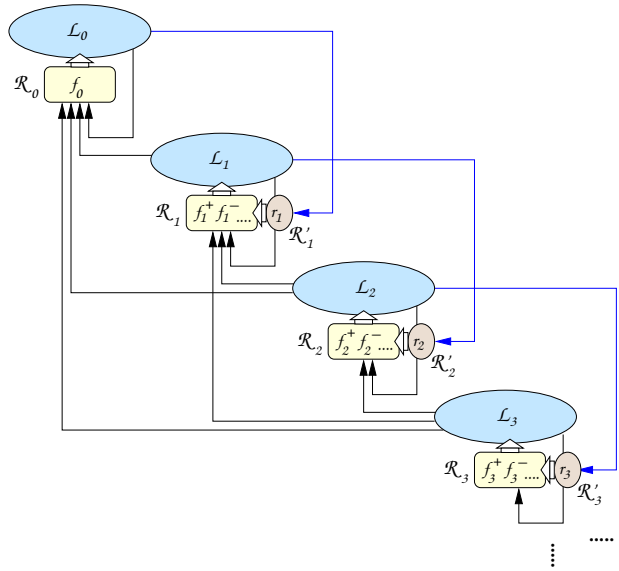}}

\caption{{\bf Beyond three levels:} General scheme to construct a HON with 
up to second order dynamics. The sets $\LL_i$ represent a hierarchy of 
types, where higher level elements correspond to groups of lower level 
elements (members). Their states are regulated by regimes $f_i$ (e.g. 
Boolean functions) taking as input states of neighbor elements at the same 
of higher levels (black arrows). The transition between the regimes is 
controlled by second order dynamics $r_i$ taking as input the members 
state at level $i-1$.}

\label{figure:HOBNwiring}

\end{figure}

The system we analyzed is an example of a Higher Order Cellular Automata 
\cite{baas05}, or even more general, a {\em higher order network} (HON). 
The first ingredient of a HON is a hierarchical structure, the idea being 
that {\it groups} of agents can act {\it together} as an entity. This 
hierarchy is modeled by creating a new agent on the next higher level, as 
illustrated in Fig. \ref{figure:HOBNwiring}. Thus we have a collection of 
sets $\LL_0, \LL_1, \ldots, \LL_n$ of agents of {\it different type}, one 
set on each level. The hierarchical structure is expressed by assigning to 
each element on level $n+1$ the set of its {\it members} on level $n$. 
Note that this is a fixed structure which will not change over time. The 
second ingredient consists again of {\it neighborhoods}. Neighbors of an 
element on level $i$ are other agents who can take influence on the state 
of the element. They can be of the same type as the element itself (that 
is, level $i$-agents) or they can be higher level agents. So the 
neighborhood of a level $i$-agent consists of subsets in levels $i$ and 
higher. The state of an element on level $i$ is regulated by the states of 
its neighbors, which serve as input for a Boolean regime. In a higher 
order network we assign {\it sets of regimes} $\R_i$ on each level, in 
other words, the system now has {\it regime states}, or {\it first order 
derived system states}. The latter name corresponds to the fact that the 
rules describe change of state. The choice of regimes is regulated by {\it 
second order rules}. Input for these second order rules can be chosen 
depending on the context of the model. With this type of wiring we create 
type-level feed-back loops containing primary interaction through direct 
state control (neighbors) and secondary interaction through regime control 
(members).

\begin{figure}

\centerline{\includegraphics[width=3.2in]{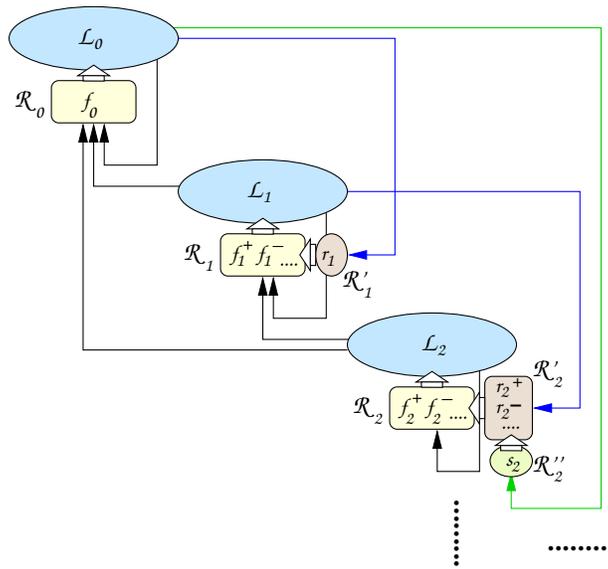}}

\caption{{\bf Third order dynamics:} Extension of the regulatory structure 
to include third order dynamics. We could also imagine a situation where 
there are more than one regime switch rule associated with the elements at 
one level, as illustrated at the third level (box $R^\prime_2$). The 
transition between these regime switchers can be controlled by a third 
order dynamics taking as input the state of elements at a lower level.}

\label{figure:HOBNwiring3} 
\end{figure}

The configuration of the regime change rules at a given time step $t$ can 
be thought of as {\it second order derived system state}: it describes the 
second order derivative of the state function. Naturally, if we allow {\it 
choice} again for these rules, we can extend this to third order. We would 
then pick another rule determining the ``change of regime switch'' and 
depending on states of certain subsets of elements (see Fig. 
\ref{figure:HOBNwiring3}). Our example concerning the cell state stops at 
the second order regulatory structure, but one can think further. For 
instance, as the chromatin strand folds within the nucleus, certain 
regions of DNA can become inaccessible to both epigenetic state modifiers 
and transcription factors \cite{grewal03}. This process does not affect 
the epigenetic state of the folded regions. However, the procedure does 
cause silencing of the genes within the folded regions. In our model, this 
would mean that the folding process has locally eliminated epigenetic 
influence on transcription. In other words, it has turned off the 
epigenetic switch between silent and accessible regimes. The epigenetic 
states are still there, but whatever they are, the transcription regime is 
silent. Thus the switching mechanism has been exchanged for a constantly 
silent one by a master-switch which depends on the folding structure.

\section{\bf Discussion}

The study of Boolean networks allows us to understand the characteristic 
features of the cell dynamics despite the great complexity of cell 
regulatory networks. A fundamental pre-requisite to achieve this goal is 
the use of ensembles of Boolean networks whose average properties are 
representative of the cell behavior. It is clear that a multi-level system 
such as the one described above can as well be encoded as an ordinary 
Boolean network. However, such a network will be a very rare realization 
within the entire set of Boolean networks with no pre-defined level 
organization. In other words, the average properties derived from the 
study of ordinary Boolean networks are not representative of a cellular 
system with its natural hierarchical structure. Higher order Boolean 
networks are therefore a necessary step to obtain network ensembles with 
pre-defined level-wise structure, a distinctive feature of cell regulatory 
networks.

Our main conclusion from a first qualitative calculation of a higher order 
model is quite striking. We can show that determining the stability of the 
cell state can be a simple or a complex problem depending on the stability 
condition for each level individually (\ref{bounds}). When the stability 
condition is satisfied for all levels we can guarantee that the system is 
stable. Similarly, when the stability condition is not satisfied for any 
level we can assure that the system is in a unstable state. In these cases 
we would say that, although the system has a second order structure, it is 
{\em simple}, i.e. its state can be determined from the analysis of its 
components independently from each other. In contrast, in between the 
simple dynamical regimes described above, there is a third regime where 
some levels do and others do not satisfy the stability condition. In this 
latter case we cannot deduce the stability of the system from the analysis 
of the stability of each single level. The system is {\em complex}, i.e. 
we are forced to consider all levels at once to determine its stability. 
This evidence indicates that the cell can be in four different states: 
{\em Simple stable} when all levels satisfy the stability condition. This 
would imply that the probability for change in any of the cells components 
is zero or converges to zero. The system is {\em complex stable} or {\em 
complex unstable} when there is at least one level that satisfies and 
another that does not satisfy the stability condition, but the system as a 
whole is stable or unstable. This could represent for example somatic 
cells in multicellular organisms with tissue regeneration (e.g., humans), 
which are epigenetically stable but may exhibit different dynamical 
behaviors at the gene and protein levels. Finally, {\em simple unstable} 
when all levels do not satisfy the stability condition. A potential 
example of this extreme case could be cancer cells, which manifest 
continuous transformations at the epigenetic, gene and protein levels.

We can further draw first rudimentary conclusions on the factors that 
influence changes between these dynamic modes based on the linear analysis 
of near-stable regimes. Our set-up allows us to weigh the contributions of 
the different cell components against each other and determine their 
comparative influence using the control structure given by the type-level 
wiring. We show that the primary factor in regulating stability is a 
dynamic mode involving all cell regulatory mechanisms (cycle $D$ in Fig. 
\ref{figure:cycles}), in particular also epigenetics. To our current 
knowledge there are so far no ensemble models in the literature which 
integrate epigenetic influence into gene expression in a systematic 
fashion which separates the different regulation mechanisms on the system 
level. In our model we can distinguish epigenetic factors from other cell 
components and account for the special role of epigenetic transcription 
regulation in a biologically sensible and accessible way.

Our approach also allows us to investigate the influence of 
``microscopic'' parameters such as neighborhood sizes, membership sizes 
and Boolean function properties. We obtain that the increase on 
neighborhood size, at any level, push the systems towards the unstable 
regime. In contrast, the increase in protein complex sizes makes the 
system more unstable. This mathematical result has important biological 
implications. It tell us that if, during the course of evolution, both the 
number of regulatory interactions and the protein complex sizes are 
increased, then the cell can remain in a nearly stable regime.

For the sake of simplicity we have focused our attention on Boolean 
models. This framework can be generalized to the case when there are more 
than two states chosen from some alphabet. More generally the states can 
take different values in algebraic groups or fields \cite{jarrah08}. We 
have freedom to choose different updating regimes as well. In the linear 
regime the form of the matrix $A$ is only determined by the topology of 
the wiring diagram, while the updating scheme just affects the actual 
values of the non-zero matrix elements.

Higher order structures are a powerful means of expressing intricate 
relations in regulatory networks of all kinds \cite{baas06}. We have shown 
here that structures of this kind are natural and adequate candidates for 
modeling biological processes. Such models are systematically more exact 
than single-level models since they formally represent patterns and types 
of regulations in the correct way and allow us to resolve the relative 
contribution of the different cellular components. They are called to play 
an even more fundamental role when addressing problems at the 
multicellular level. We hope this work motivates further efforts towards 
the annotation of cell regulatory networks, making an explicit distinction 
between the different cellular components, their level organization, and 
feedback regulation.

\section*{Appendix: Partial derivatives of $\Lambda$}

To compute the derivatives of $\Lambda$ with respect to $A_\E$, $A_\G$ 
and $A_\X$, we take into account that 

\begin{equation}
\frac{\partial P}{\partial a_{ij}} =
(-1)^{i+j} P_{ij}(\Lambda) + 
\frac{\partial P}{\partial\Lambda} 
\frac{\partial\Lambda}{\partial a_{ij}}\ ,
\end{equation}

\noindent where $P_{ij}$ is the characteristic polynomial associated with 
the minor of $A$ after removing line $i$ and column $j$. Furthermore

\begin{equation}
\frac{\partial P}{\partial D} = -1 + 
\frac{\partial P}{\partial\Lambda} 
\frac{\partial\Lambda}{\partial D}\ ,
\end{equation}

\noindent From these equations it follows that

\begin{equation}
\frac{\partial P}{\partial a_{ij}} = 
(-1)^{i+j} P_{ij} \frac{\partial\Lambda}{\partial D}\ ,
\end{equation}

\noindent Now, since $A$ is irreducible and non-negative, the largest 
eigenvalue of $A$ is larger or equal than the largest eigenvalue of any 
submatrix of $A$ (Frobenius Theorem \cite{gantmacher00,debreu53}). The 
latter result implies that $P_{ij}(\Lambda)\geq0$. To show that $P_{ij}<1$ 
we need to inspect the precise form of $P_{ij}$. For $i=j=1$ we obtain

\begin{equation}
P_{11}(\Lambda) = \Lambda(\Lambda^3 - (A_\G+A_\X)\Lambda^2
+ A_\G A_\X \Lambda - C\ .
\end{equation}

\noindent Since $\Lambda\geq A_G$ (Frobenius Theorem) we have that

\begin{eqnarray}
P_{11}(\Lambda) &\leq& \Lambda(\Lambda^3 - (A_\G+A_\X)\Lambda A_\G
+ A_\G A_\X \Lambda - C)\nonumber\\
&\leq& \Lambda(\Lambda^3 - A_\G^2\Lambda - C)\nonumber\\
&\leq& \Lambda^4\ .
\end{eqnarray}

\noindent For $\Lambda<1$ we finally obtain that $P_{11}(\Lambda)\leq1$ 
and therefore

\begin{equation}
\frac{\partial P}{\partial A_\E} =
\frac{\partial P}{\partial a_{11}} \leq 
\frac{\partial P}{\partial D}\ .
\end{equation}

\noindent Following a similar analysis we obtain that

\begin{equation}
\frac{\partial P}{\partial A_\G} =
\frac{\partial P}{\partial a_{22}} \leq
\frac{\partial P}{\partial D}\ .
\end{equation}

\begin{equation}
\frac{\partial P}{\partial A_\X} =
\frac{\partial P}{\partial a_{33}} \leq
\frac{\partial P}{\partial D}\ .
\end{equation}

On the other hand, for the derivatives with respect $B$ and $D$ we 
obtain

\begin{eqnarray}\label{dBdDdC}
\frac{\del \Lambda}{\del B} &=& \Lambda (\Lambda-A_\E)
\frac{\del \Lambda}{\del D}\\
\frac{\del \Lambda}{\del C} &=& \Lambda (\Lambda-A_\X)
\frac{\del \Lambda}{\del D}\ .
\end{eqnarray}

\noindent Now again $\Lambda$ is larger than the largest eigenvalue of any 
submatrix (Frobenius Theorem \cite{gantmacher00}) and thus it is larger 
than any matrix element. In particular $\Lambda\geq a_{11}=A_\E$ and 
$\Lambda\geq a_{33}=A_\X$. Under the assumption $\Lambda<1$, these results 
then imply that $0\leq\Lambda(\Lambda-A_\E)\leq1$ and 
$0\leq\Lambda(\Lambda-A_\E)\leq 1$, and therefore

\begin{eqnarray}\label{dBdDdCineq}
\frac{\del \Lambda}{\del B} &\leq&
\frac{\del \Lambda}{\del D}\\
\frac{\del \Lambda}{\del C} &\leq&
\frac{\del \Lambda}{\del D}\ .
\end{eqnarray}

\section*{References}


\end{document}